\documentclass{epsconf}
\usepackage{graphicx,dcolumn,bm,sidecap}
\usepackage[scriptsize,bf]{caption}  
\usepackage{wrapfig}
\usepackage{amsmath}
\title{High flux expansion divertor studies in NSTX}
\author{V. A. Soukhanovskii$^1$,
R. Maingi$^2$,
R. E. Bell$^3$,
D. A. Gates$^3$,
R. Kaita$^3$,
H. W. Kugel$^3$,\\
B. P. LeBlanc$^3$,
R. Maqueda$^4$,
J. E. Menard$^3$,
D. Mueller$^3$,
S. F. Paul$^3$,
R. Raman$^5$,\\
A. L. Roquemore$^3$
}
\institute{$^1$ Lawrence Livermore National Laboratory, Livermore, CA, USA\\
$^2$ Oak Ridge National Laboratory, Oak Ridge, TN, USA\\
$^3$ Princeton Plasma Physics Laboratory, Princeton, NJ, USA\\
$^4$ Nova Photonics, Princeton, NJ, USA\\
$^5$ University of Washington, Seattle, WA, USA\\
}
\begin{document}
\maketitle
{\bf{Introduction}}
Projections for high-performance H-mode scenarios in spherical torus (ST)-based devices assume low electron collisionality for increased efficiency of the neutral beam current drive.
At lower collisionality (lower density), the mitigation techniques based on
induced divertor volumetric power and momentum losses may not be capable of reducing heat and material erosion to acceptable levels in a compact ST divertor.
Divertor geometry can also be used to reduce high peak heat and particle fluxes by flaring a scrape-off layer (SOL) flux tube at the divertor plate, and by optimizing the angle at which the flux tube intersects the 
divertor  plate, or reduce heat flow to the divertor by increasing the length of the flux tube.
The recently proposed advanced divertor concepts \cite{9586707, valanju:056110} take advantage of these geometry effects.
\begin{wrapfigure}[24]{r}{83mm} 
\setlength\abovecaptionskip{-1pt}
\vspace{-4mm}
\resizebox{80mm}{!}{\includegraphics*[angle=0]{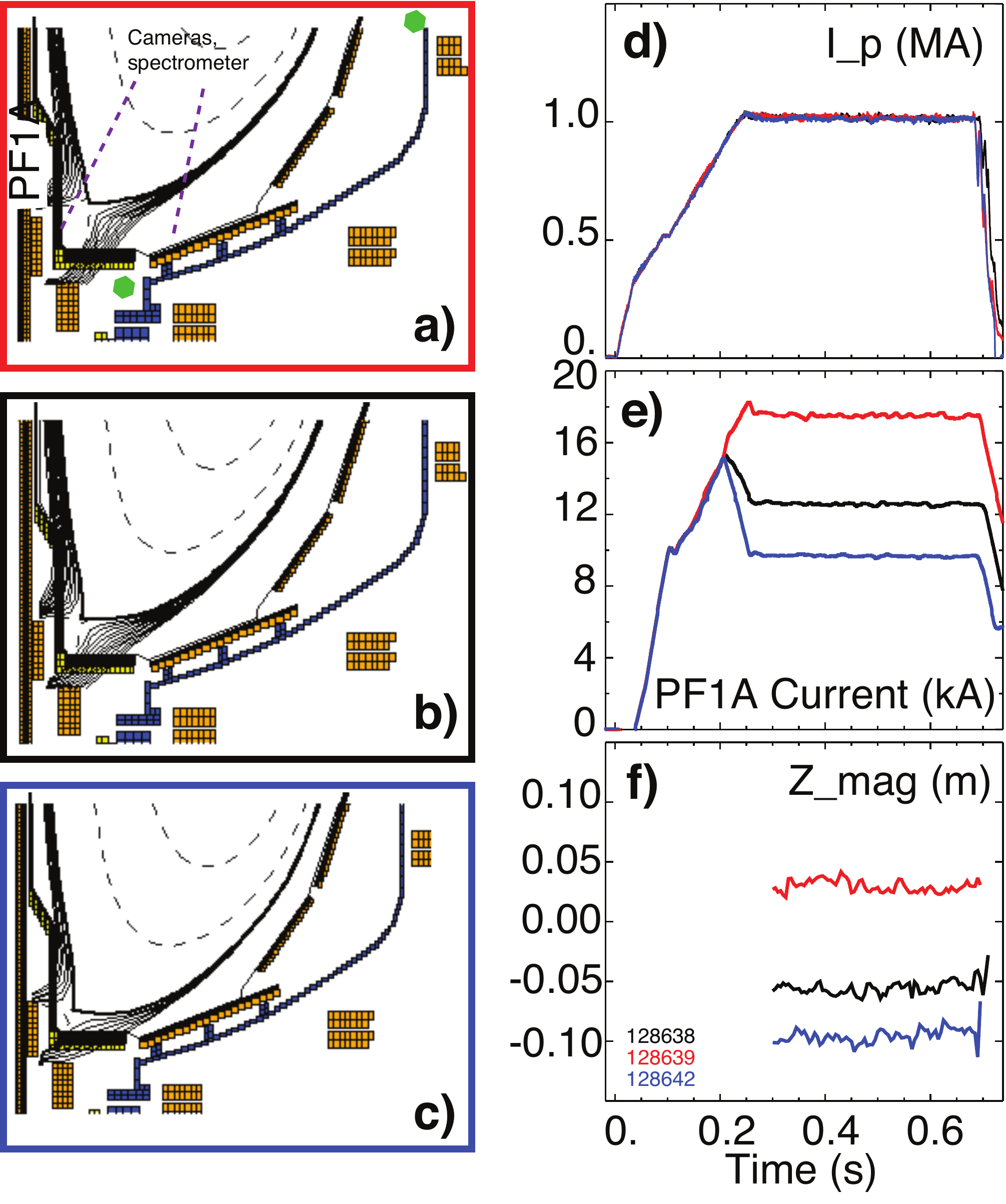}}
\caption{\it{Divertor equilibria for $h_{X} = 18-22$ cm (a), $h_{X} = 8-12$ cm (b), $h_{X} = 3-6$ cm (c).
The shown flux surfaces are separated by 1 mm radial distance in the midplane, and show a full extent of the divertor SOL for each configuration.
The associated (color-coded with same color) traces of plasma current (d), divertor PF1A coil current (e), and plasma magnetic axis vertical position (f). }}
\label{f:one}
\end{wrapfigure}
In a high triangularity ST plasma configuration, the magnetic flux expansion at the divertor strike point (SP) is inherently high, leading to a reduction of
heat and particle fluxes and a facilitated access to the outer SP detachment, as has been demonstrated  recently in NSTX \cite{vas-pop09}. 
The natural synergy of the highly-shaped high-performance ST plasmas with beneficial divertor properties motivated a further systematic study of the high flux expansion divertor.

The National Spherical Torus Experiment (NSTX) is a mid-sized device with the aspect ratio 
$A=1.3-1.5$ \cite{074010841128}.
In NSTX, the graphite tile divertor has an open horizontal plate geometry.
The divertor magnetic configuration geometry was systematically changed in an experiment by either (i) changing the distance between the lower divertor X-point and the divertor plate  (X-point height $h_{X}$), or by (ii) keeping the X-point height constant and increasing the outer SP radius. 
An initial analysis of the former experiment is presented below.
Since in the divertor the poloidal field $B_{\theta}$ strength is proportional to $h_X$, the X-point height variation changed the divertor plasma wetted area due to variations in a SOL flux expansion, or more generally, area expansion.
The flux expansion factor is defined as 
$f_{exp} = \lambda_q^{SP} /  \lambda_q^{MP} \simeq (B_{\theta}/B_{tot})^{MP} / (B_{\theta}/B_{tot})^{SP}$,
where $B_{tot}$ is the total magnetic field at the strike point $R_{SP}$ 
and midplane $R_{MP}$ major radii, and $\lambda_q$ is the SOL parallel heat flux width. The SOL area expansion factor is  $f_{exp} \;R_{MP}/R_{SP}$.\\
\noindent
{\bf{Experiment}}
The experiment was carried out in a lower single null configuration, in 1~MA, 6~MW NBI-heated H-mode plasma discharges, fueled by deuterium from a high field side gas injector.
Shown in Fig.~\ref{f:one} (a)-(c) are three representative configurations with X-point heights $h_{X} = 18-22, 9-12$, and 5-6 cm. 
For the divertor measurements, infrared thermography, spectrally filtered cameras, a multi-channel UV-visible spectrometer, and Penning and micro-ion neutral pressure gauges were used, as described in detail in \cite{vas-pop09}.
Steady-state configurations with different $h_X$ were obtained on a shot to shot basis 
by shifting the plasma up or down, as shown in Fig.~\ref{f:one} (f).
The steady-state shift was accomplished by controlled variations of the divertor PF1A coil current within 10-18 kA (Fig.~\ref{f:one} (e)), while
other plasma shaping parameters were kept within a reasonably small range:
elongation 2.20-2.40, triangularity 0.7-0.85, and midplane radial distance between the primary and secondary separatrices ($dr_{sep}$)  5 - 12 mm.\\

\noindent
{\bf{Results and discussion}}
Large aspect ratio tokamak divertor experiments have provided insights into the 
scrape-off layer heat  flux variations with X-point height  \cite{6979968,6487522,6262610,6908766}.
In this section we summarize the NSTX observations and discuss them in the context of the mentioned results. 

\begin{wrapfigure}[22]{r}{63mm}
\setlength\abovecaptionskip{-2pt}
\vspace{-5mm}
\resizebox{60mm}{!}{\includegraphics*[angle=0]{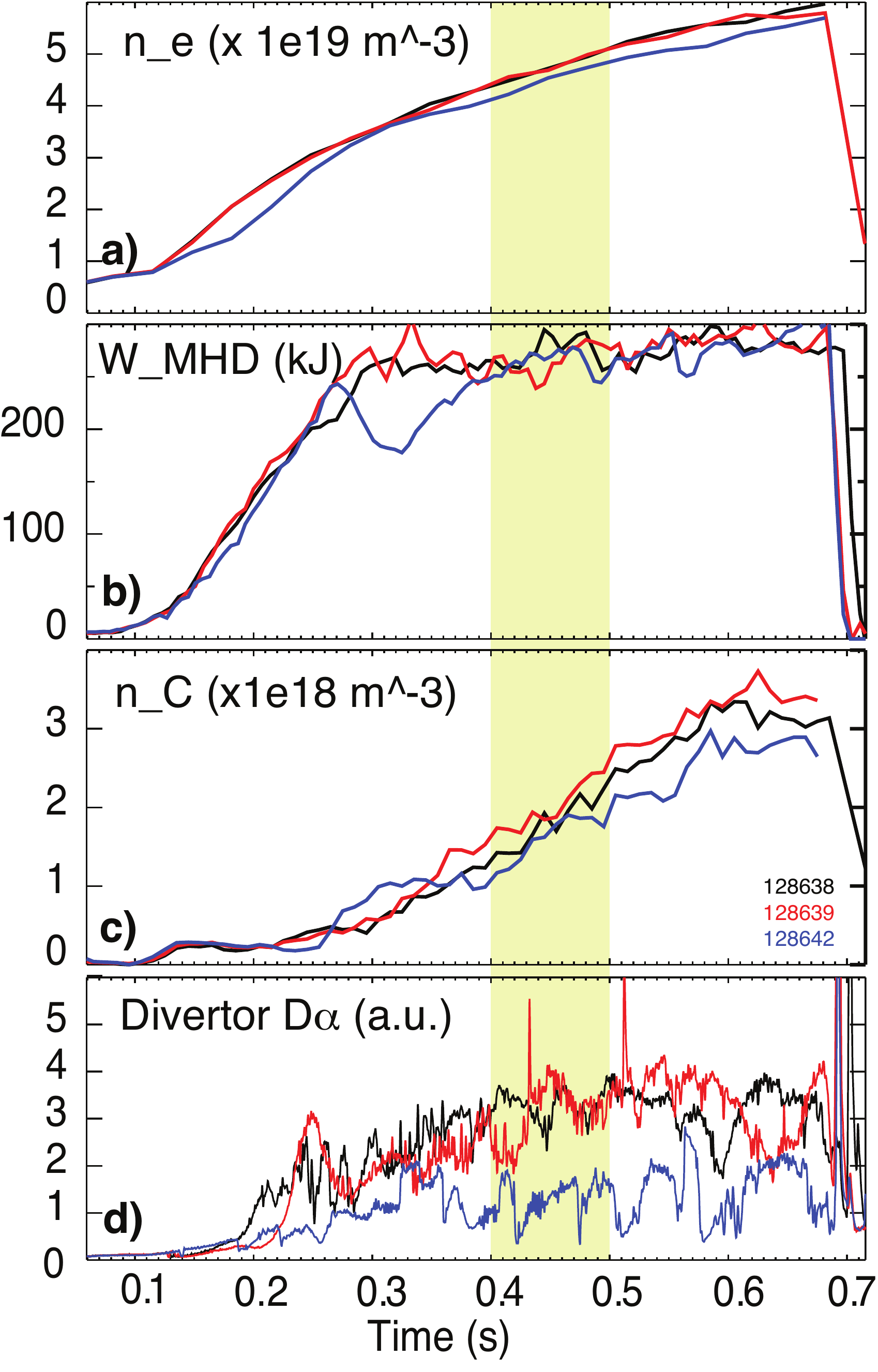}}
\caption{\it{Time traces of core plasma parameters for the configurations shown in Fig.~\ref{f:one}
in same colors (a) plasma density $\bar{n}_e$, (b) Plasma stored energy, (c) core carbon density, and  (d)  total lower divertor D$_{\alpha}$ intensity.
}}
\label{f:two}
\end{wrapfigure}
Core plasma parameters varied within 10-15 \%  over the range of the X-point height  scan.
As shown in Fig.~\ref{f:two} for the medium, low, and high $h_{X}$, the plasma stored energy 
remained nearly constant, while the electron density and core carbon density modestly varied.
A portion of the variation was due to instrumental effects, as the NSTX Thomson scattering ($T_e, n_e$) and charge-echange recombination spectroscopy ($n_C, T_i$) systems probed the plasma at $Z=0.0$ m,
while the actual plasma center was shifted by 5-10 cm up or down.
The small-ELM H-mode regime was maintained regardless of the divertor configuration, 
with occasional  large type I ELMs. 
The midplane (upstream) SOL $T_e$ and $n_e$ did not appear to change, remaining in the range 30-50 eV, and $3-8 \times 10^{18}$ m$^{-3}$, respectively.

\begin{wrapfigure}[29]{r}{63mm} 
\setlength\abovecaptionskip{-2pt}
\vspace{-8.5mm}
\resizebox{60mm}{!}{\includegraphics*[angle=0]{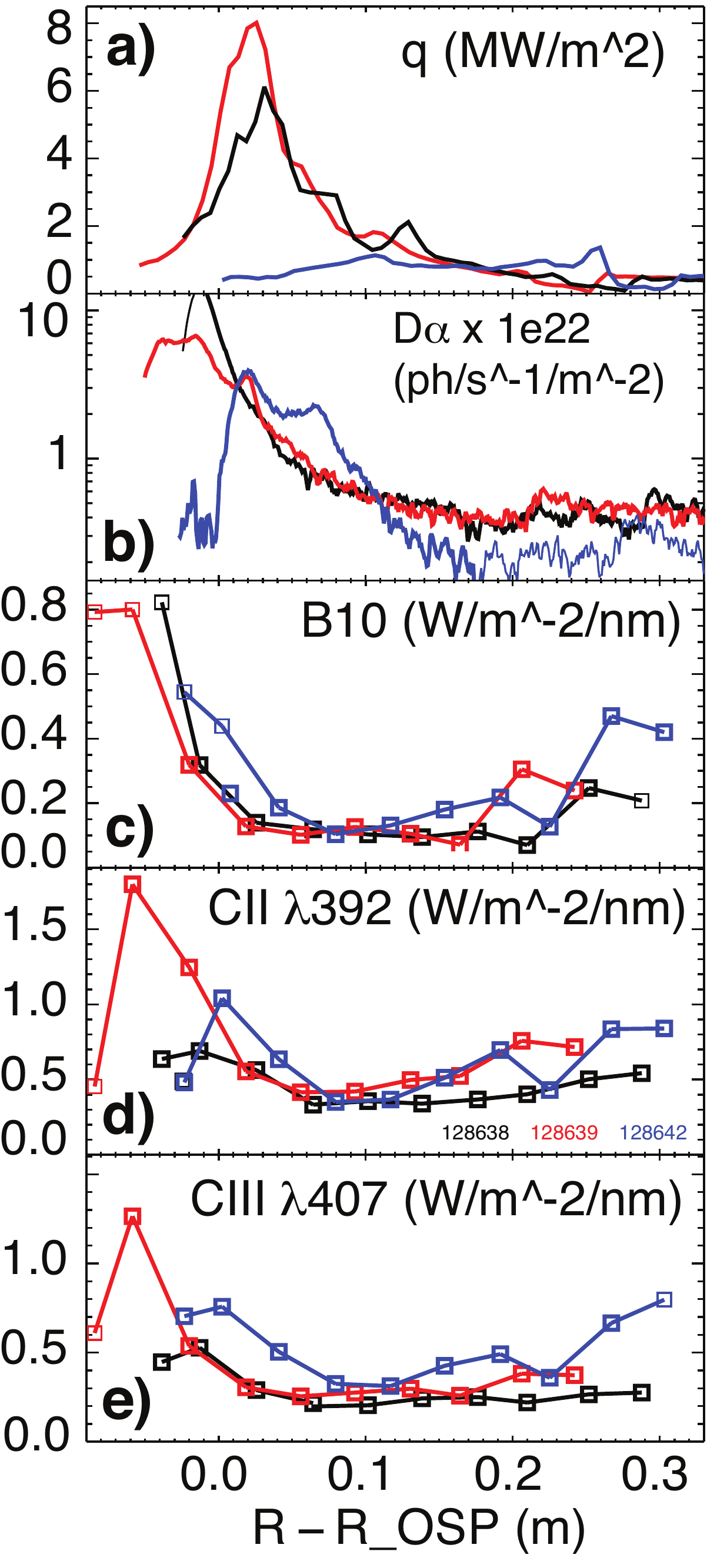}}
\caption{\it{ Lower divertor profiles at 0.422-0.480 ms: (a) heat flux, and brightness profiles (b) D$_{\alpha}$, (c) DI Balmer B10 ($\lambda=380$ nm), (d) C II ($\lambda = 397$ nm), and (e) C III ($\lambda = 407$ nm).}}
\label{f:three}
\end{wrapfigure}

As expected, $h_X$ variations had a profound effect on divertor profiles.
Overall, the divertor measurements indicated that as $h_{X}$ was decreased, the conduction-dominated high-recycling SOL developed a higher parallel $T_e$ gradient, as the divertor $T_e$ apparently decreased, and $n_e$ increased. 
The increase in divertor $n_e$ and recombination rate suggested that at the lowest $h_{X}$, the outer SP region was nearly detached.
The profiles shown in Fig.~\ref{f:three} were measured between ELMs during the time interval indicated in Fig.~\ref{f:two}, unless indicated otherwise. 
The heat flux profiles showed a drastic reduction in peak heat flux from $q_{pk}$=7-8 MW/m$^2$ in the high $h_{X}$ case, down to 1-2 MW/m$^2$ in the lower  $h_{X}$ case.
The width of the profiles $\lambda_q$ also consistently increased with flux expansion.
Because of the strong $B_{\theta}/B_{tot}$ variation across the separatrix and the heat diffusion into the private flux region, heat flux profiles do not peak at the separatrix \cite{4167597}.
This was apparent in all $h_X$ cases. However, at the lowest $h_X$ $q_{pk}$ was several cm away from the separatrix location, suggesting a partial SP detachment.
Outside of the divertor SOL, the heat flux did not change, and remained at a level of 0.5-1 MW/m$^2$.
The $D_{\alpha}$ brightness was higher in the outer SP region at lower $h_{X}$, although
the $D_{\alpha}$ profiles were more difficult to interpret since the X-point region was dominated by emission and could only be partially resolved from the outer SP region (Fig.~\ref{f:three} (b)).
Spatially resolved ultraviolet spectra showed the presence of the high-$n$ Balmer series lines.
Upper levels of these lines are populated by three-body recombination.
Volume recombination took place in all $h_{X}$ cases in the inner divertor, however, the Balmer $n=10-2$  (B10) line emission intensity increased, and the emission region expanded with lowering the X-point, and spread from the inner divertor and X-point region to the outer SP region, as shown in Fig.~\ref{f:two} (c).
This was also consistent with the observed outer SP D$_{\alpha}$ intensity increase, as part of the increase could be due to recombination.
Preliminary analysis of the Balmer series Stark broadening indicated an increase in divertor density to $n_e \leq 3 \times 10^{20}$ m$^{-3}$.
Emission from C II and C III also increased in the X-point - outer SP region in the lower $h_{X}$ case (Fig.~\ref{f:three} (d, e)).

\begin{wrapfigure}[27]{r}{63mm} 
\setlength\abovecaptionskip{-2pt}
\resizebox{60mm}{!}{\includegraphics*[angle=0]{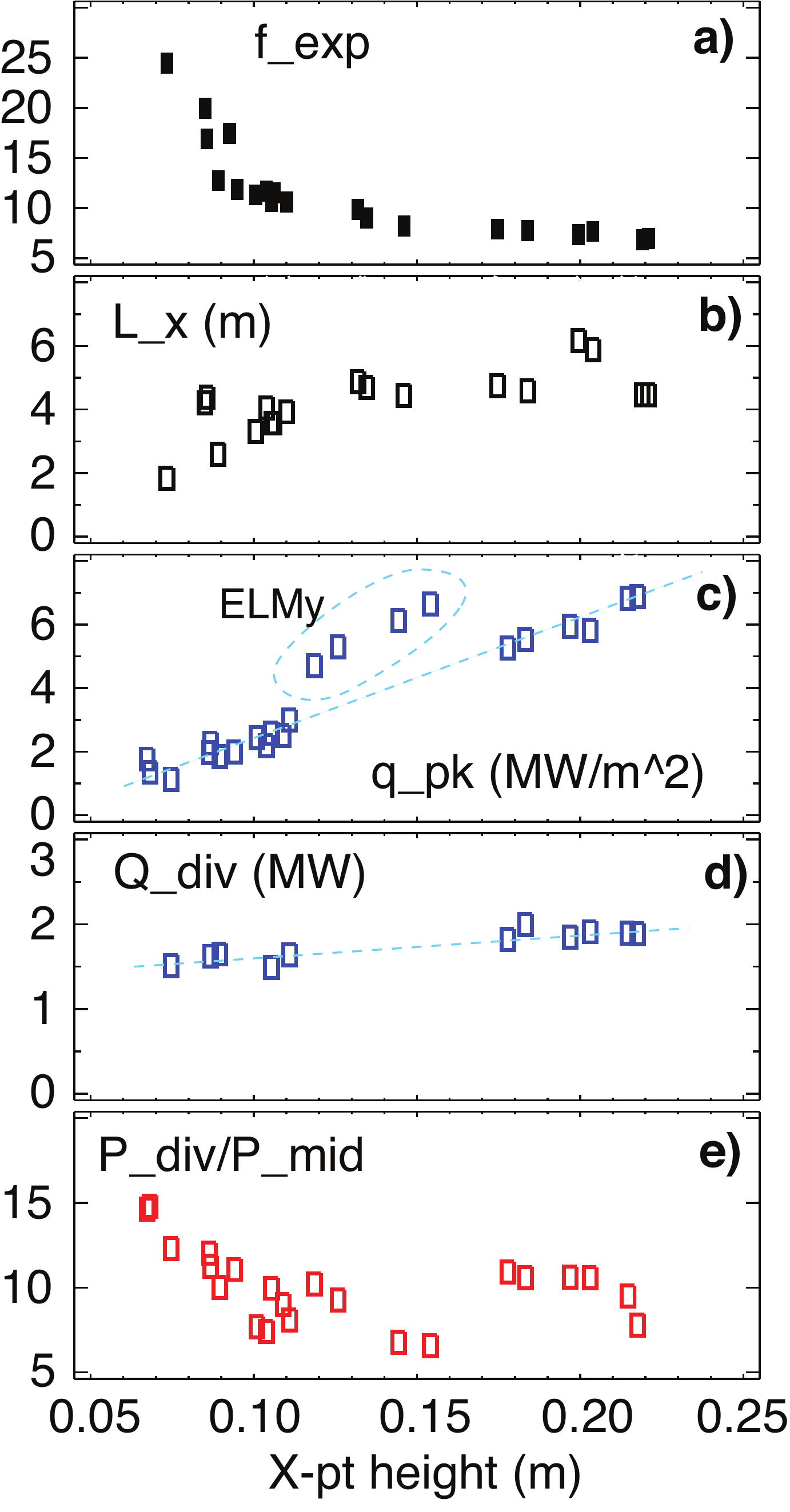}}
\caption{\it{Divertor geometry and plasma parameters as functions of $h_X$: (a) flux expansion, (b) X-point connection length, (c) peak divertor heat flux, (d) peak divertor heating power, (c) divertor deuterium compression.}}
\label{f:four}
\end{wrapfigure}

Divertor geometry changes resulting from the $h_{X}$ scan are summarized in Fig.~\ref{f:four} (a) and (b). 
The flux expansion factor decreased from about 26-28 to 6 as $h_{X}$ was changed from 7 to 23 cm.
The outer SP moved radially outward by 10 cm over the $h_X$ scan.
The corresponding area expansion factor decreased from 5.5 (at $h_X = 7$ cm) to 1.8 (at $h_X = 22$ cm).
The connection length $L_c$ between the midplane and divertor target  remained about $15\pm 2$ m, while the X-point to target connection length $L_x$ (both calculated at $\psi_N = 1.001$) changed from 2 to 5-6 m.
The field line incidence angle $\gamma$ changed from 1-2 to 4-5 degrees in the $h_X$ scan.
Divertor $q_{pk}$ was nearly proportional to $h_{X}$, except in one discharge where heat flux was measured over frequent medium size ELMs, as indicated in Fig.~\ref{f:four} (c).
The peak divertor power $Q_{div}$ changed weakly between 1.5-2 MW (Fig.~\ref{f:four} (d)), suggesting that some losses were 
incurred due to increased radiated power at lower $h_X$.

With an orthogonal divertor target orientation (i.e., when flux surfaces are perpendicular to divertor targets, as was the case), the SOL heat flux $q_{\parallel}$ is reduced due to the power dissipated in the divertor, conduction and convection losses, and the geometry factor $B_{\theta}/B_{tot}$ \cite{s00}.
In a conduction-dominated SOL, the heat loss due to private flux region diffusion is proportional to the X-point connection length $L_x$, while the conduction losses are proportional to $L_X^{4/7}$ \cite{s00}.
Additionally, the $f_{exp}$ variation with $h_X$ was much stronger than the $L_X$ variation (Fig.~\ref{f:four} (a) and (b)), suggesting that besides the dissipative losses, the flux expansion effect is the dominant effect for heat flux reduction in this divertor configuration.
When the peak heat flux $q_{pk}$ is mapped to the midplane, the obtained $q_{\parallel} = q_{pk}/\sin \gamma$ is in the range 50-100 MW/m$^2$, being lower at the lower $h_X$.
While this result is in qualitative agreement with the large aspect ratio tokamak divertor studies \cite{6066501},
the further clarification of the private flux region diffusion effect in NSTX without accurate accounting of the divertor dissipative losses is difficult. 
The data suggested that at the lowest $h_X$ the divertor radiated power losses were non-negligible, and the strike point was close to detachment.
At the medium and high $h_X$, the assumption of the conduction-dominated flux-limited outer SOL regime
was supported by this and other NSTX studies 
\cite{05098861864,072110609441,072110609413,Ahn2009421,10407123}.

Variations in the X-point height had measurable effects on divertor particle handling.
Lowering $h_{X}$ apparently  led to an increased opacity to neutrals.
Divertor pressure measurements indicated a nearly linear increase of $P_{div}$ with decreasing $h_{X}$.
Shown in Fig.~\ref{f:three} (e) is the divertor compression factor $P_{div} / P_{mid}$, a ratio of divertor and midplane neutral pressures. The divertor
 compression trend and the deuterium emission profiles in Fig.~\ref{f:three} suggested that 
neutral entrapment ("plugging efficiency") was improved with decreasing $h_X$.
The lowest $h_X$ case will be analyzed further, as the poloidal length of the divertor leg was approaching
several ion gyroradii, while the connection length $L_X$ was marginal for efficient ion recombination to occur in the divertor.

In summary, divertor configuration studies by means of the X-point height variation in NSTX showed that  
in a high triangularity configuration, the poloidal flux expansion could be varied from 6 to 26.
The flux expansion factor appeared to play the dominant role in peak divertor heat flux reduction from 7-8 MW/m$^2$ to 1-2 MW/m$^2$.
When the X-point height was reduced to several cm, the divertor $T_e$ decreased and $n_e$ increased, the radiated power and recombination rate increased, suggesting that the strike point region was approaching the detachment state.
The X-point variation studies also suggested that a further clarification of the relative roles of various divertor heat reduction and loss mechanisms would be better accomplished experimentally by maintaining a medium X-point height while scanning the flux expansion and outer strike point radius.\\
 
\noindent
{\bf{Acknowledgments}} We thank the entire NSTX Team for technical, computer and engineering support, as well as for 
plasma, NBI and diagnostic operations. This work was performed under the auspices of the U.S. Department of Energy under Contracts DE-AC52-07NA27344,  DE-AC02-76CH03073, DE-AC05-00OR22725, W-7405-ENG-36.
\let\oldthebibliography=\thebibliography
  \let\endoldthebibliography=\endthebibliography
  \renewenvironment{thebibliography}[1]{%
    \begin{oldthebibliography}{#1}%
      \setlength{\parskip}{0ex}%
      \setlength{\itemsep}{0ex}%
  }%
  {%
    \end{oldthebibliography}%
  }

\end{document}